\documentclass[conference]{IEEEtran}
\IEEEoverridecommandlockouts
\usepackage{cite}
\usepackage{amsmath,amssymb,amsfonts}
\usepackage{algorithmic}
\usepackage{graphicx}
\usepackage{fancyhdr}
\usepackage{textcomp}
\usepackage{multirow}
\usepackage{xcolor}
\usepackage[english]{babel}
\usepackage{subcaption}
\usepackage{multirow}
\usepackage{textcomp}
\usepackage{orcidlink}
\usepackage{textcomp}
\usepackage{pgfplots}
\pgfplotsset{width=8cm,compat=1.9}
\def\BibTeX{{\rm B\kern-.05em{\sc i\kern-.025em b}\kern-.08em
    T\kern-.1667em\lower.7ex\hbox{E}\kern-.125emX}}
    
\fancypagestyle{firstpage}{
    \fancyhf{} 
    \fancyfoot[L]{tinyML Foundation EMEA Innovation Forum 2024} 
}

\begin{document}

\title{Enhancing TinyML Security: Study of Adversarial Attack Transferability \\

}

\author{\IEEEauthorblockN{Parin Shah \orcidlink{0009-0004-2232-1638}, Yuvaraj Govindarajulu \orcidlink{0000-0002-4247-4410}, Pavan Kulkarni \orcidlink{0000-0002-8458-6795}, Manojkumar Parmar \orcidlink{0000-0002-1183-4399}}
\IEEEauthorblockA{\textit{AIShield, Bosch Global Software Technologies} \\
\textit{Bangalore, India}\\
\{parin.shah , govindarajulu.yuvaraj\}@bosch.com }
}

\maketitle
\thispagestyle{firstpage}
\begin{abstract}
The recent strides in artificial intelligence (AI) and machine learning (ML) have propelled the rise of TinyML, a paradigm enabling AI computations at the edge without dependence on cloud connections. While TinyML offers real-time data analysis and swift responses critical for diverse applications, its devices' intrinsic resource limitations expose them to security risks. This research delves into the adversarial vulnerabilities of AI models on resource-constrained embedded hardware, with a focus on Model Extraction and Evasion Attacks. Our findings reveal that adversarial attacks from powerful host machines could be transferred to smaller, less secure devices like ESP32 and Raspberry Pi. This illustrates that adversarial attacks could be extended to tiny devices, underscoring vulnerabilities, and emphasizing the necessity for reinforced security measures in TinyML deployments. This exploration enhances the comprehension of security challenges in TinyML and offers insights for safeguarding sensitive data and ensuring device dependability in AI-powered edge computing settings.
\end{abstract}


\section{Introduction}
Artificial intelligence (AI) simulates human intelligence in machines, enabling them to reason and acquire information similarly to people. This field of computer science develops machines capable of tasks requiring human intelligence, like understanding language, recognizing images, and making quick decisions.
\\
TinyML, short for Tiny Machine Learning, is a subfield of AI specializing in deploying machine learning models on small, resource-constrained devices like microcontrollers and sensors. These devices face limitations in computing power, memory, and energy, making traditional machine learning models challenging to run. TinyML overcomes these constraints by designing small, efficient models that operate on low-power devices.\\
Emerging technology, TinyML \cite{RAY20221595} has the potential to revolutionize our interactions with the world. The development of small, low-power devices capable of advanced machine learning tasks like image and speech recognition opens doors in various industries, from healthcare to industrial automation. Applications include monitoring vital signs in remote locations, detecting and responding to equipment malfunctions, and providing real-time feedback for personal fitness. However, like any technology, TinyML poses security risks.
\par Deploying TinyML systems in sensitive environments like healthcare and industrial control raises significant security concerns. These systems often handle sensitive data and control critical infrastructure. Additionally, their resource constraints and limited computational power make them vulnerable to attacks.

\par The rapid progress of AI has introduced remarkable breakthroughs but it also raises concerns about adversarial attacks \cite{10.1007/978-3-031-44137-0_32} \cite{govindarajulu2023targeted}. These attacks manipulate AI systems through deceptive input, compromising decision-making. The paper \cite{Biggio_2018} delves into the advancements in adversarial AI, shedding light on the vulnerability of AI  models 
especially deep networks, to intentional input perturbations designed to exploit weaknesses in AI systems. These refer to malicious actions that exploit vulnerabilities in AI systems. Such attacks can manifest in various forms and can target different aspects of an AI system, including data, models, or infrastructure. To systematically identify potential security attacks and defense mechanisms for a given use case, threat model, which defines the capabilities and goals of the attacker under realistic assumptions, is required. Adversarial attacks are categorized into Black Box Attacks, Grey Box Attacks, and White Box Attacks, based on the extent of the attacker's knowledge regarding the targeted AI system's inner workings and components.
\begin{enumerate}
    \item Black Box Attack: Black Box AI attacks \cite{Biggio_2018} refer to malicious actions or activities that exploit vulnerabilities in AI systems without requiring knowledge of the internal workings of the system. These types of attacks are particularly challenging to defend against because they do not rely on a known vulnerability or weakness in the system. 
    \item Grey Box Attack: Grey Box attacks \cite{10.1007/978-3-642-40994-3_25} involve exploiting vulnerabilities in AI systems where the attacker has partial knowledge of the internal workings of the system. These types of attack can involve combining different pieces of information that the attacker may have access to, such as knowledge of the model architecture or training data. These attacks are different from black-box attacks, where the attacker has no information about the system.
    \item White Box Attack: In a white-box attack \cite{bhambri2020survey}, the adversary has complete knowledge about the target model, including learned weights, parameters, and sometimes labeled training data. The common strategy involves modeling the distribution from the known weights to generate perturbed inputs that can strategically breach the model boundaries. This approach leverages the attacker's in-depth understanding of the model's inner workings, making white-box attacks advanced and effective in compromising AI systems.
\end{enumerate}

\section{Related Work}

The authors \cite{REN2020346} provide an overview of the current state of adversarial machine learning research. They review the progress made on adversarial attacks and defenses over the past decade and discuss the limitations of current defense techniques. They also present an overview of the most recent and promising research directions in this field, such as the development of more robust models, the use of adversarial training, and the use of more advanced methods for detecting and mitigating adversarial attacks.
\par TinyML refers to the field of machine learning applied to small, low-power devices such as sensors, wearables and Internet of Things (IoT) devices \cite{9586232}. Current progress in TinyML includes advances in techniques for compression, quantization, and energy-efficient computing to enable machine learning models to run on small devices. Research challenges include improving model accuracy, reducing power consumption, and addressing the limitations of current hardware. The future roadmap for TinyML includes continued research in these areas, as well as the development of new technologies and architectures to further enable the deployment of machine learning on small devices.
\par 
There is a growing body of research on benchmarking TinyML systems, with studies focusing on various aspects such as evaluation methods, datasets, and metrics. Benchmarking \cite{unknown} these systems can be challenging due to the diversity of devices and use cases, as well as the lack of standardized evaluation methods and datasets. However, benchmarking is important for comparing the performance of different TinyML systems and for guiding the development of new ones. Some potential directions for benchmarking TinyML systems include developing standardized evaluation methods, creating benchmark datasets that reflect real-world scenarios, and creating metrics that take into account not only accuracy but also factors such as power consumption and memory usage.
\par In paper \cite{Shafique_2021}, the authors explore the realm of Edge AI systems, examining challenges and presenting techniques to improve performance, energy efficiency, reliability, and security. They introduce a cross-layer framework integrating cutting-edge methods to enhance energy efficiency and robustness in Edge AI. Additionally, they discuss advancements and challenges in neuromorphic computing, particularly focusing on Spiking Neural Networks (SNNs). The work aims to provide researchers and practitioners with insights into the evolving landscape of Edge AI systems and the strategies driving their progress.

\subsection{This Paper}\label{AA}
\par The aim of this research is to examine the  vulnerabilities of AI models against adversarial attacks on resource-constrained embedded hardware. We evaluate the transferability of adversarial attacks, on embedded hardware using Model Extraction and Evasion Attacks. The attacks initiates on a host machine, following which we proceeded to deploy the models onto two different hardware devices, namely the ESP32 and Raspberry Pi. Our result demonstrate that adversarial attacks are transferable from powerful host machines to smaller, less secure tiny devices. This highlights a concerning vulnerability on the deployed embedded hardware against adversarial attacks, necessitating enhanced security measures for AI models deployed on tiny systems.

\par 

\section{Methodology}
\subsection{Architectural Details of Embedded Hardware Used in the Experiments}
\begin{figure*}[t]
\centering     
\begin{subfigure}[b]{0.24\textwidth}
\includegraphics[width=\textwidth]{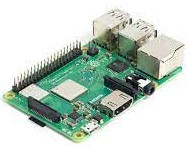}
\caption{Raspberry Pi 3B+}
\end{subfigure}
\hspace{0.2\textwidth}
\begin{subfigure}[b]{0.24\textwidth}
\includegraphics[width=\textwidth]{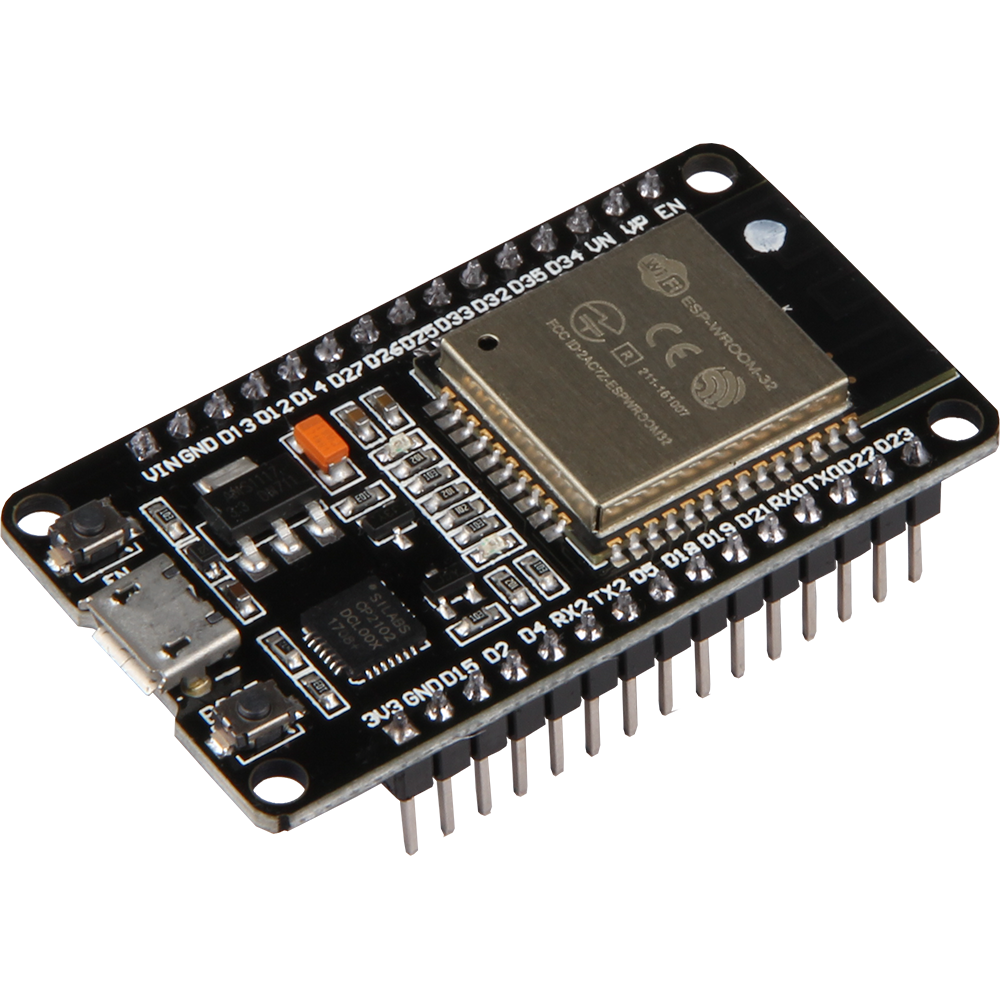}
\caption{ESP 32}
\end{subfigure}
\caption{Visualizing the hardware used in the experiments}
\label{fig:common_caption}
\end{figure*}
\subsubsection{Raspberry Pi}

The Raspberry Pi 3B+ \cite{raspberrypiRaspberryModel}, as shown in Fig. 1 (a) stands out as a compelling platform for exploring and implementing AI applications. Its cost-effectiveness makes it an accessible gateway for individuals and organizations to experiment with AI, fostering innovation and development. The 64 bit quad-core processor and diverse I/O ports provide the necessary horsepower for tackling essential AI tasks, while compatibility with renowned frameworks like TensorFlow and PyTorch empowers users to leverage pre-trained models and development tools. This synergy between affordability, capability, and ease of use positions the Raspberry Pi 3B+ as a valuable asset for AI education, research, and hobbyist endeavors. As AI continues to mature and permeate our daily lives, the Raspberry Pi 3B+ is poised to remain a relevant and popular platform for those seeking to engage with this transformative technology.
\subsubsection{ESP32 Wroom Dev KitC}
The ESP32 \cite{espressifDevKitsEspressif}, illustrated in Fig. 1 (b), with its dual-core processor and rich wireless connectivity, carves a niche in the realm of edge AI. This versatile microcontroller boasts impressive processing power (240 MHz) to handle complex AI tasks efficiently. Built-in Wi-Fi, Bluetooth, and BLE unlock seamless communication within IoT networks, making it ideal for AI applications like smart home systems and sensor data analysis. Notably, its low power consumption makes it the perfect choice for battery-powered devices, enabling portable and sustainable AI solutions. 

\par Empowering its AI capabilities, the ESP32 readily handles large datasets and is compatible with frameworks like TensorFlow Lite \cite{tensorflowTensorFlowLite}, allowing on-device AI computations without relying on the cloud. This combination of processing power, wireless connectivity, and efficient power management makes the ESP32 an adaptable tool for diverse AI projects, ranging from IoT applications and home automation to sensor networks. As edge AI continues to evolve, the ESP32 is poised to remain a prominent player, offering a robust and versatile platform for bringing intelligence to the very edge of computing.
\renewcommand{\arraystretch}{1.5}
\begin{table}[t]
\centering
\caption{Overview on Hardware Specifications}
\label{tab:my_label}
\begin{tabular}{|p{2.2cm}|p{2.2cm}|p{2.5cm}|}
\hline
 & \textbf{ESP 32} & \textbf{Raspberry Pi}  \\
\hline
\textbf{CPU} & Xtensa LX6 &ARM Cortex-A53 \\
\hline
\textbf{Memory} & 520 KB SRAM &1GB LPDDR2 SDRAM \\
\hline
\textbf{Storage} & -&  SD Card \\
\hline
\textbf{OS} & -&Raspbian OS \\
\hline
\textbf{Optimized Framework} & TFlite Micro &TFLite \\
\hline
\end{tabular}
\end{table}

\par Table I summarizes the specifications and special features, along with the optimized AI frameworks supported by the hardware used during the experiment.

\subsection{Adversarial Attacks Employed for Subsequent Experiments} 
\subsubsection{Model Extraction Attack}
 Model extraction attack \cite{tramèr2016stealing} is a method of stealing the knowledge of the pre-trained model (targeted model) and transferring it to a surrogate model. It is achieved by carefully curating queries to exploit the targeted model in any one of the settings specified earlier. Since we artificially generate queries for the targeted model, the targeted model predicts the output corresponding to the curated input. Thus, we achieve a surrogate data set that can be used to train your custom model, which on completion of training mimics the functionality of the targeted model. Model extraction attack \cite{lekkala2021emerging} starts out with the attacker querying the targeted model with a large number of inputs in order to get as much information about the model's functionality as possible. The attacker can get insights into the model's decision-making process, such as decision boundaries, feature importance, and model architecture, by analyzing the model's outputs. 
 \par Once the attacker accumulates enough knowledge about the targeted model, they can utilise it to train a substitute model that mimics the original model's behaviour. This substitute model can then be used to generate effective adversarial instances for attacking the original model. The model extraction attack has the advantage of not requiring access to the model's architecture or parameters, making it a successful attack even when the model's specifics are not publicly known. The effectiveness of this attack, however, is dependent on the quality and quantity of data collected by the attacker, as well as the capacity to train a substitute model that accurately matches the behaviour of the targeted model. Consider a trained model \((f)\) or victim model (\textit{V}) deployed on the device, an adversary can pass an attack vector as input \((x)\) to obtain a prediction \((y)\) on input feature vectors through the non-secure communication channel \((NSComm)\) . The adversary will eventually be able to reconstruct a Learned Labeled Data Set \((LLDS)\). The attacker can then train a replication model \((f_e)\) or surrogate model (\textit{S}) on \(LLDS\) to approximate \((f)\) without prior knowledge about its parameters.
\subsubsection{Evasion Attack}
The Evasion attack \cite{7090993} is a type of adversarial attack that seeks to manipulate a AI models by altering the input data so that the model miss-classifies it. In other words, the attacker creates a changed input, known as an adversarial example, that is identical to the original input but can lead to an inaccurate prediction by the model. The goal of evasion attack is to determine the smallest change to the original input that will result in a miss-classification. The attacker can accomplish this by maximizing the discrepancy between the predicted output of the original input and the intended output using optimization techniques. This difference is often evaluated with a loss function that measures the cost of miss-classification.


\begin{figure}[t]
    \centering
    \includegraphics[scale = 0.5]{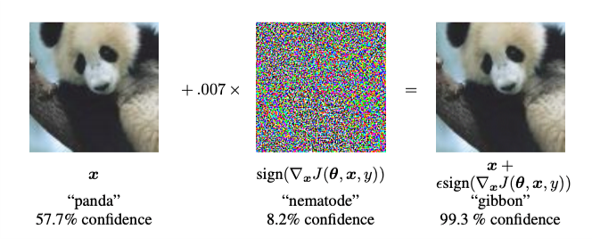}
    \caption{An example of Evasion Attack \cite{goodfellow2015explaining}}
    \label{fig:my_label}
\end{figure}
\subsubsection{Fast Gradient Sign Method -- FGSM}
Fast Gradient Sign Method \cite{goodfellow2015explaining} (FGSM) is a type of White Box Attack that aims to generate adversarial examples by computing the gradient of the loss function with respect to the input data and using it to generate perturbations to the input. This method performs by applying a small perturbation to each feature of the input data in the direction of the gradient of the loss function with respect to that feature. The size of the perturbation is controlled by a small value known as the attack's strength, which determines the degree of the attack. The FGSM's key advantage is its simplicity and efficiency, as it determines the gradient with a single forward and backward run over the model. However, the attack's success is determined by the attack strength and the vulnerability of the targeted model to small perturbations.
In Fig. 2, we have an example where the starting image represents a panda. The attacker proceeds to introduce subtle perturbations or distortions to the original image. Consequently, the model exhibits high confidence of 99.3 $\%$ in misclassifying this image as a Gibbon.

\par The adversarial examples \(adv_x\) are generated by following formula:

\begin{center}
    \(adv_x = x + \epsilon*sign(\nabla_{{x}} J(\theta, x, y))\)
\end{center}

where 
\begin{enumerate}
    \item \(adv_x\) = Adversarial Data
    \item \(x\) = Original Input Data
    \item \(y\) = Labels corresponding to Input Data (x)
    \item \(\theta\) = Original Model 
    \item \(\epsilon\) = Scalar value to control the magnitude of perturbation
    \item \(J(\theta, x,y)\) = Loss Function of the Model
     \item \(\nabla_x J(\theta,x,y)\) = Gradient of the loss function with respect to input (x)
    \item  \(sign(\nabla_{\mathbf{x}} J(\theta, x, y))\) = Sign of the Gradient, which is either -1 or 1)
\end{enumerate} 
\subsection{Hardware Attacks and Security}
\label{sec:HWAttacks}

\par Robust hardware security requires safeguarding embedded firmware, data, and overall system functionality. This becomes especially crucial when protecting sensitive information like cryptographic keys or personal data. Unhindered access to firmware poses a grave threat, allowing attackers to delve into the program and potentially uncover vulnerabilities, bypass licensing, and override software restrictions. This breach could lead to replicating custom algorithms or deploying cloned hardware. Even in open-source environments, ensuring code authenticity is paramount to prevent malicious firmware insertion. Furthermore, denial-of-service (DoS) attacks \cite{boraten2018mitigation, fang2013robustness} pose a significant threat to crucial systems like environmental monitoring (e.g., gas, fire) and security apparatuses like intrusion detection alarms or surveillance cameras. While implementing stringent security measures can introduce complexity, it's essential to maintain the resilience and reliability of these systems, ensuring their uninterrupted and dependable operation. IoT, or smart devices, have increased the demand for security. Hackers find connected devices very attractive because they can access them remotely. Protocol vulnerabilities offer an angle of attack through connectivity. Fig. 3 illustrates a successful attack scenario where a single compromised device can jeopardize the integrity of an entire network.
\par The ever-evolving landscape of hardware security demands constant vigilance against a diverse range of attacks. Each category of attack presents unique challenges, necessitating the development of specialized defense mechanisms to effectively counter them
\begin{itemize}
    \item Software Attacks: Software attacks exploit vulnerabilities in the code, such as bugs or protocol weaknesses, and can be executed remotely without physical access to the device. These attacks, often leveraging untrusted pieces of code, pose a significant threat, with interception or usurpation of communication channels being common tactics. Software attacks, being widespread and relatively inexpensive, represent the majority of cases, emphasizing the need for robust defenses against code-based vulnerabilities.
    \item Hardware Attacks: Hardware attacks \cite{electronics6030052} require physical access to the device, adding an additional layer of complexity to the threat landscape. The most apparent hardware attack involves exploiting the debug port, particularly if it lacks adequate protection. However, hardware attacks, in general, are sophisticated endeavors that can incur substantial costs. These attacks require specialized materials and electronics engineering skills. They are further categorized into noninvasive attacks, conducted at the board or chip level without causing device destruction, and invasive attacks, which occur at the device-silicon level and often involve package destruction.
\begin{figure}[t]
    \centering
    \includegraphics[scale=0.35]{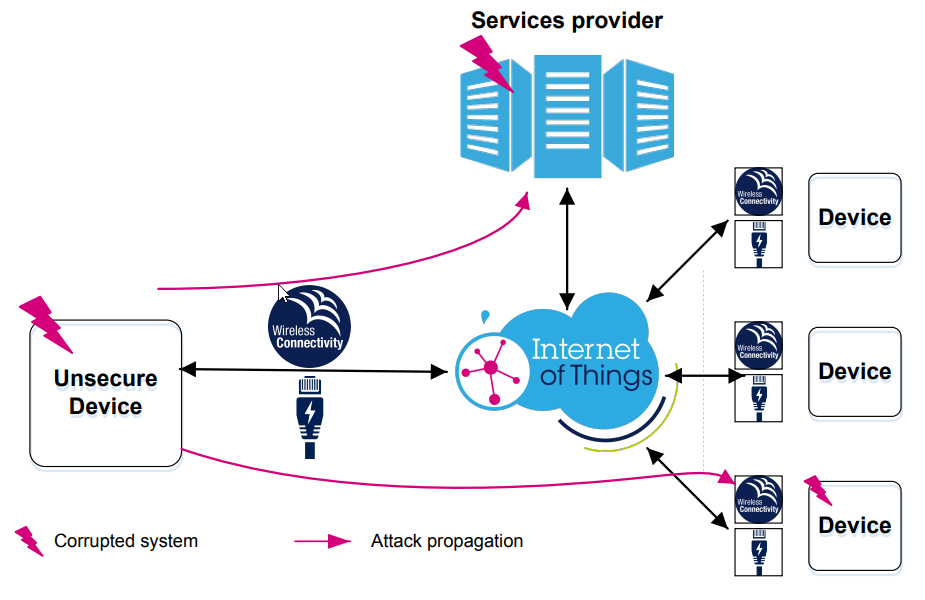}
    \caption{Impact of a Single Compromised Device on Network Integrity \cite{key}}
    \label{fig:enter-label}
\end{figure}
\end{itemize}
\par The security measures rely on hardware mechanisms activated based on careful configuration using option bytes or dynamic orchestration by hardware components. These safeguards are essential in various critical areas.
\begin{itemize}
    \item Memory Protection: It involves implementing measures to prevent unauthorized access, manipulation, or exploitation of crucial program instructions and data segments. This is achieved by employing hardware mechanisms such as Memory Management Units (MMUs) and access control lists, ensuring that only authorized entities can interact with specific areas of the system's memory.
    \item Software Isolation: It focuses on strengthening the internal architecture of the system to prevent potential breaches and maintain the integrity of individual processes. 
    \item Interface Protection: It aims to mitigate the risks associated with external attacks that exploit vulnerable entry points to gain unauthorized access to the device. This involves implementing secure communication protocols, input validation mechanisms, and applying strict access controls to external interfaces, reducing the attack surface and strengthening the overall security posture.
    \item System Monitoring: It strengthens the system against external attacks employing Intrusion Detection Systems (IDS), real-time monitoring tools, and anomaly detection algorithms. These continuously observe system behavior, promptly identify potential threats, and initiate appropriate responses to safeguard system integrity against tampering or anomalous activities
\end{itemize}
\par Hardware security measures \cite{electronics12214507,ravi2004security}, while effective in safeguarding physical components through encryption, secure boot processes, and tamper resistance, are insufficient to defend against AI attacks. The evolving nature and complexities of artificial intelligence threats render hardware-centric approaches inadequate. AI attacks exploit vulnerabilities in the software layer, targeting algorithms, data inputs, and model architectures. Examples include adversarial examples, data poisoning, model inversion, and model stealing. Unlike conventional security threats, AI attacks manipulate AI systems at their core, subtly altering inputs or exploiting data biases to produce incorrect or malicious outcomes. Furthermore, AI systems are dynamic and adaptive, continuously learning and evolving based on new data and experiences. This dynamic nature poses significant challenges for hardware-based security solutions, which may struggle to keep pace with the rapid changes in AI models and techniques. 
\par As a result securing AI models deployed on hardware is crucial given the evolving cybersecurity threats. Although hardware measures such as encryption and tamper resistance provide foundational protection, they alone cannot fully address the dynamic nature of AI attacks. These attacks exploit vulnerabilities within AI algorithms and data, necessitating a comprehensive strategy that combines hardware and software defenses. Software-based measures such as adversarial training, model verification, and data integrity checks are essential for fortifying AI models against emerging threats like adversarial examples and data poisoning. Essentially, the integration of both hardware and software security measures is crucial for nurturing trust and robustness in AI-driven technologies across a multitude of applications and industries.

\section{Experimental Setup and Attack Preparation}

\par In this section, we present a comprehensive overview of the experimental results, focusing on the evaluation of AI models' performance on the host machine—an x64-based processor system with 32 GB RAM and a 6 GB NVIDIA RTx A3000 GPU. The experiments involved executing adversarial attacks on two publicly available datasets: MNIST \cite{deng2012mnist}, an image classification dataset, and Gesture Recognition, which serves as an illustration of a time series dataset. The MNIST dataset comprises handwritten digits ranging from 0 to 9, while the Gesture Recognition \cite{githubGitHubStefanspissMagicWandTFLiteESP32MPU6050} dataset consists of three distinct classes: Ring, Wing, and Slope. Our objective was to assess the susceptibility of the AI models to malicious attacks by conducting two types of attacks: Model Extraction attack and Model Evasion attack. The outcomes of these attacks are elaborated upon in subsequent sections, providing a detailed account of the results.



\subsection{Results for MNIST Dataset}

In this section, we present the experimental results derived from our analysis of the MNIST dataset , focusing on the outcomes of model extraction and evasion attacks. Simultaneously, we assess the real-world performance of our AI model by deploying them on the the resource-constrained tiny intelligence RPi. Our experiment encompasses an in-depth examination of model accuracy, susceptibility to adversarial attacks, and their implications on RPi. This section offers comprehensive understanding of adversarial attack delivering an in-depth discussion of our observations and findings.

\subsubsection{Model Extraction Attack}
In this experiment, a Convolutional Neural Network (CNN) trained on the MNIST dataset, referred to as our victim model (\textit{V}), was utilized, achieving a 99.3 $\%$ accuracy. For the model extraction attack, attack queries were generated. These attack queries are passed through \textit{V} to generate labels, which are then utilized to train the  surrogate model (\textit{S}) model, mimicking the behavior of \textit{V}. Fig. 4 illustrates the attack vectors employed during the training of the \textit{S}.  Impressively, despite having only 14K parameters, \textit{S} achieved a notable 90 $\%$ accuracy. 
\par Table II provides a summary of the outcomes from the model extraction attack, presenting a comparison between the accuracy and model size of the \textit{V} and the \textit{S}.


\renewcommand{\arraystretch}{1.5}
\begin{table}[t]
  \centering
  \label{tab:my_label}
    \caption{Result of MNIST Dataset Model Extraction Attack on Host}
  \begin{tabular}{|c|c|c|}
    \hline
  \textbf{Model} &  Victim Model (\textit{V}) & Surrogate Model (\textit{S})\\ 
    \hline
{\textbf{Accuracy ($\%$)}} & 99.3 & 90  \\ \hline
\textbf{Model Format} & \multicolumn{2}{|c|}{H5}  \\ \hline \cline{2-3} 
\textbf{Model Size (MB)} & 4.66 & 1.67 \\ \hline

  \end{tabular}
\end{table}

\begin{figure}[t]
    \centering
    \includegraphics[scale =0.3]{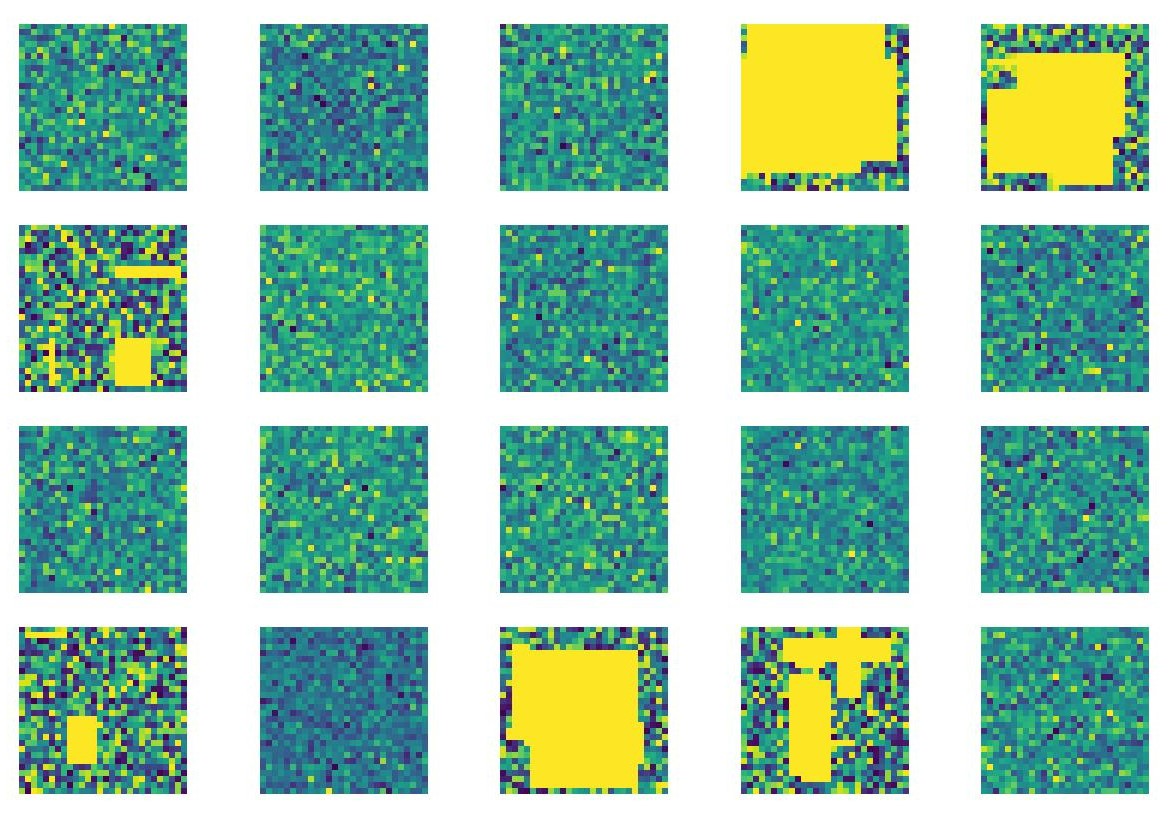}
    \caption{Visualizing Attack Vectors used in the training of surrogate  model}
    \label{fig:my_label}
\end{figure}

\subsubsection{Evasion Attack}
After executing the model extraction attack, we proceeded to conduct an evasion attack on \textit{V} using the FGSM. In this attack, we crafted attack vectors by introducing imperceptible perturbations to the input dataset. These subtle alterations, invisible to the human eye, have the potential to induce misclassification by the model. The magnitude of these perturbations, referred to as attack strength, directly impacts the extent to which the model's robustness is compromised.
\par Fig. 5 visually demonstrates the effects of different attack strengths, illustrating how they can lead to the miss-classification of classes in the original dataset. To assess the effectiveness of our evasion attack, we quantified changes in the model's classification accuracy on the perturbed dataset. Fig. 6 graphically depicts the relationship between attack strength and efficacy, highlighting that higher attack strength resulted in an increased number of misclassified classes. Notably, with ($\epsilon$) = 0.31, we flipped 67.9 $\%$ of the original class. This emphasizes the significant impact and implications of evasion attacks on AI models.

\begin{figure}[t]
    \centering
    \includegraphics[scale=0.2]{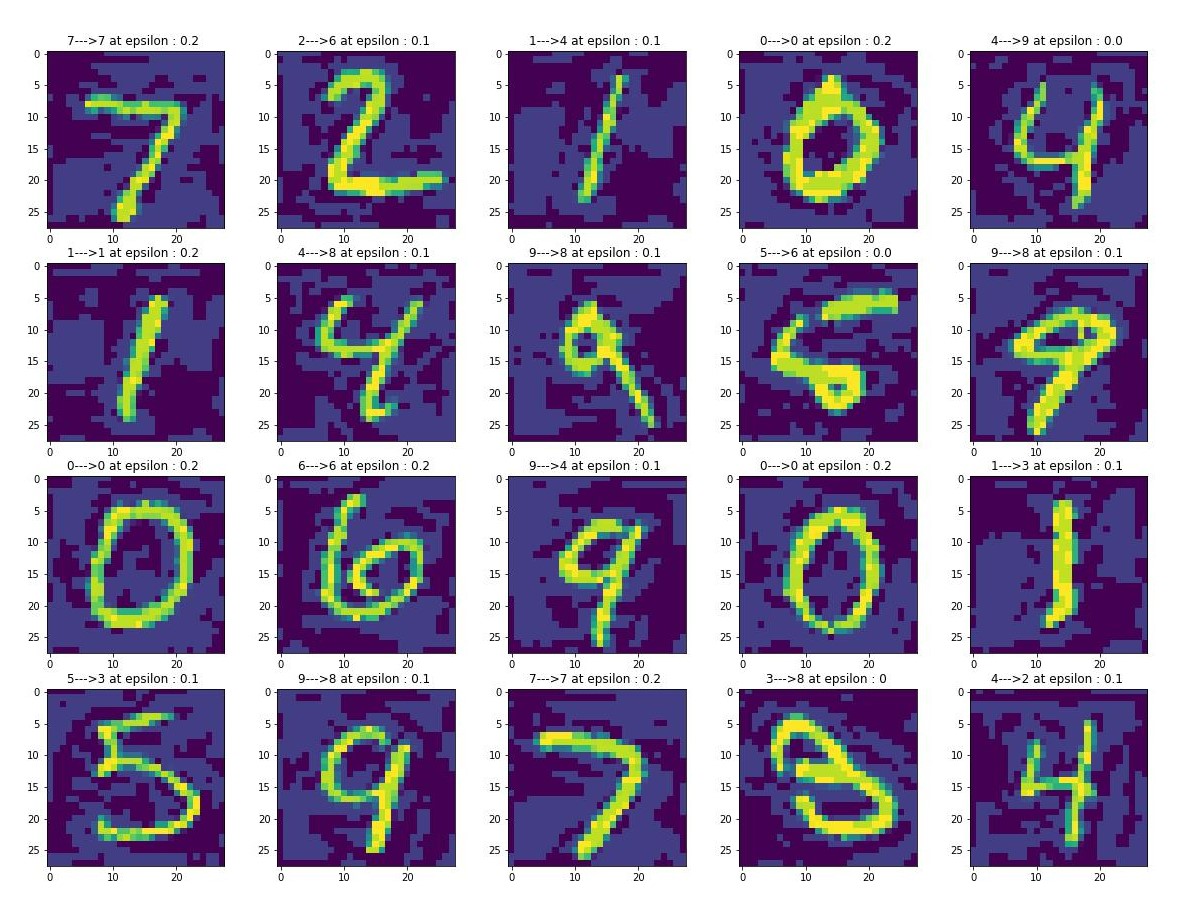}
    \caption{Original class being missclassified to different class}
    \label{fig:my_label}
\end{figure}





\begin{figure}[t]

\begin{tikzpicture}
\begin{axis}
[
    xlabel={Attack Strength \((\epsilon)\)},
    ylabel={Efficacy (\%)},
    xmin=0, xmax=0.4,
    ymin=0, ymax=100,
    xtick={0,0.05,0.1,0.15,0.2,0.25,0.3,0.35, 0.4},
    ytick={0,20,40,60,80,100,120},
    legend pos=north west,
    ymajorgrids=true,
    grid style=dashed,
]
\addplot[
    color=blue,
    mark=square,
    ]
    coordinates {
    (0.01,0.25)(0.03,4.2)(0.08,16.3)(0.1,24.6)(0.2,58.56)(0.3,67.6)(0.31,67.39)(0.35,67.6)(0.4,67.5)
    };
    \end{axis}
\end{tikzpicture}
\caption{Effect of Attack Strength on Original Dataset} 
\label{fig:M1}
\end{figure}
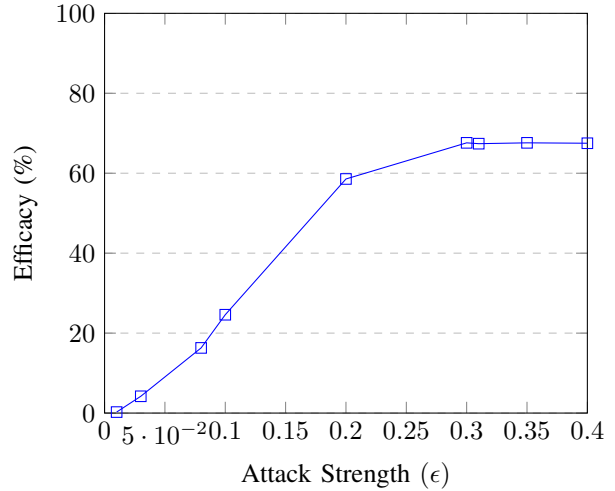


\subsection{Results of Gesture Recognition Dataset}

In this section, we will detail the outcomes of our experiments conducted on the Gesture Recognition Dataset, encompassing three distinct classes: Ring, Wing, and Slope. As a component of the experiment, we executed two adversarial attacks, namely model extraction and evasion. Subsequent to the successful completion of these experiments on the host system, we proceeded to deploy the model on ESP32, aiming to assess the vulnerabilities of the AI model on the resource-constrained platform. This holds particular significance, considering the prevalent use of AI models on devices with limited resources, necessitating their resilience against potential adversarial attacks. This section provides a thorough comprehension of adversarial attacks, presenting an in-depth discussion of our observations and findings.

\subsubsection{Model Extraction}

A Convolutional Neural Network (CNN)-based architecture was selected as the \textit{V}, attaining an accuracy of 92.37 $\%$. For the execution of the model extraction attack, we generated a set of attack queries directed to the \textit{V} to acquire its predictions.  Fig. 7 depicts these attack queries utilized for training the \textit{S}. By leveraging these queries and the responses from \textit{V}, we successfully trained a new \textit{S} capable of replicating \textit{V}'s behavior. Remarkably, despite possessing a mere 38K parameters, the \textit{S} attained an accuracy of 69.23 $\%$. Table III offers a comprehensive overview of the model extraction attack's results, presenting a comparative analysis of both \textit{V} and \textit{S} in terms of their accuracy and size.

\renewcommand{\arraystretch}{1.5}
\begin{table}[t]
  \centering
  \label{tab:my_label}
    \caption{Result of Gesture Recognition Model Extraction Attack on Host}
  \begin{tabular}{|c|c|c|}
    \hline
  \textbf{Model} &  Victim Model (\textit{V}) & Surrogate Model (\textit{S})\\ 
    \hline
{\textbf{Accuracy ($\%$)}} &  92.37 & 69.23  \\ \hline
\textbf{Model Format} & \multicolumn{2}{|c|}{H5}  \\ \hline \cline{2-3} 
\textbf{Model Size (MB)} &  0.173 & 0.173   \\ \hline

  \end{tabular}
\end{table}

\begin{figure}[t]
    \centering
    \includegraphics[scale=0.4]{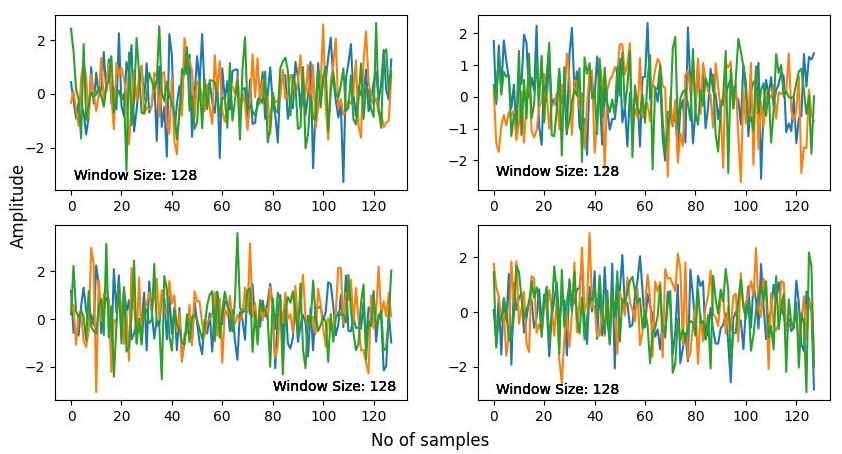}
    \caption{Attack Vectors used in Training Surrogate Model}
    \label{fig:my_label}
\end{figure}


\subsubsection{Evasion Attack}
In this section, we delve into the execution of an evasion attack on the \textit{V} model following the model extraction phase, utilizing the FGSM. Throughout the evasion attack, we crafted attack vectors by introducing imperceptible perturbations to the input dataset—undetectable to the human eye yet capable of inducing misclassifications by the model. The magnitude of these perturbations, referred to as Attack Strength, governs the extent to which the model's robustness is compromised. 
To measure the effectiveness of our evasion attack, we evaluated variations in the model's classification accuracy on the perturbed dataset. Fig. 8 presents a graph illustrating the relationship between Attack Strength and Efficacy. As depicted, an escalation in Attack Strength correlates with an increased incidence of misclassified classes.
\par Fig. 8 illustrates that with a rise in Attack Strength, there is a corresponding augmentation in the percentage of misclassified data. At an Attack Strength of ($\epsilon$) = 0.9, the evasion attack exhibited high effectiveness, resulting in a misclassification rate of 95.76 $\%$. This outcome underscores the susceptibility of the AI Model employed in the \textit{V} to adversarial attacks.
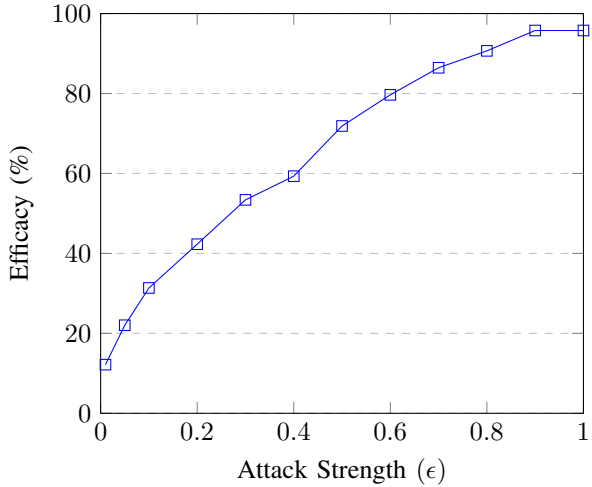
\begin{figure}[t]
    \begin{tikzpicture}

\begin{axis}
[
    xlabel={Attack Strength \((\epsilon)\)},
    ylabel={Efficacy (\%)},
    xmin=0, xmax=1.0,
    ymin=0, ymax=100,
    xtick={0,0.2,0.4,0.6,0.8,1.0,1.2},
    ytick={0,20,40,60,80,100,120},
    legend pos=north west,
    ymajorgrids=true,
    grid style=dashed,
]
\addplot[
    color=blue,
    mark=square,
    ]
    coordinates {(0.01,12.17)(0.05,22.03)(0.1,31.35)(0.2,42.3)(0.3,53.39)(0.4,59.32)(0.5,71.87)(0.6,79.67)(0.7,86.44)(0.8,90.67)(0.9,95.76)(1,95.76)
    
    };
    
\end{axis}
\end{tikzpicture}
\caption{Effect of Attack Strength on Gesture Recognition Original dataset} \label{fig:M1}
\end{figure}

\section{Experimental Results on RPi $\&$ ESP32}
Subsequent to our experiments on the host device and the ensuing outcomes, our experiment extended to evaluate the performance of the models across diverse hardware platforms. Specifically, deploying the models on resource-constrained devices such as the RPi and ESP32. The primary objective was to scrutinize the models' vulnerability to adversarial attacks concerning the hardware platform of deployment. Our thorough analysis revealed a consistent accuracy across various hardware platforms, with no noticeable deviation. Upon meticulous examination of the experimental outcomes, we concluded that the susceptibility of AI models to adversarial attacks is not limited to any specific hardware platform.
\subsection{Implementing MNIST Dataset on RPi}
In this section, our primary objective is to conduct an exhaustive analysis of the susceptibility exhibited by AI models deployed on the RPi, coupled with a meticulous summarization of the observed performance metrics. To achieve this, we systematically executed methodical experiments involving both model extraction and evasion attacks, employing the MNIST Dataset as our experimental foundation.
\subsubsection{Model Extraction}
After the successful execution of the model extraction attack on the host device, our aim was to deploy the obtained model on RPi platforms. The objective was to specifically evaluate the model's susceptibility on the resource constrained platform. Surprisingly, our observations revealed that the performance of the extracted model did not experience significant degradation in comparison to its performance on the host device. The model maintained a high accuracy of 90 $\%$, aligning with the accuracy achieved on the host device. Table IV presents a summary of the outcomes obtained from the model extraction attack, furnishing information about the deployed model format on RPi. Additionally, it includes a comparative analysis of accuracy and model size between the \textit{V} and \textit{S}.
\par This result highlights the importance of evaluating the security measures and robustness of models, even when deployed on resource-constrained hardware devices like the RPi. 

\renewcommand{\arraystretch}{1.5}
\begin{table}[t]
  \centering
  \label{tab:my_label}
    \caption{Result of MNIST Dataset Model Extraction Attack on Raspberry Pi}
  \begin{tabular}{|c|c|c|}
    \hline
  \textbf{Model} &  Victim Model (\textit{V}) & Surrogate Model (\textit{S})\\ 
    \hline
{\textbf{Accuracy ($\%$)}} & 99.3 & 90  \\ \hline
\textbf{Model Format} & \multicolumn{2}{|c|}{TFLite}  \\ \hline \cline{2-3} 
\textbf{Model Size (MB)} & 1.53 & 0.548 \\ \hline

  \end{tabular}
\end{table}
\subsubsection{Evasion Attack}

In this section, our focus shifts to evaluating the model's robustness through evasion attacks targeted at the \textit{V}. The primary objective is to assess vulnerabilities on the RPi platform. The outcomes and insights derived from these experiments are systematically presented and detailed in Table V. This table serves as a comprehensive repository, offering an expansive overview of the model's performance across a spectrum of attack intensities. The detailed analysis encompasses the model's behavior under varying degrees of attack, providing a nuanced understanding of its resilience and susceptibility in the face of evasion tactics on the RPi.

\renewcommand{\arraystretch}{1.5}
\begin{table}[t]
\centering
\caption{Result of Attack Strength vs Efficacy for MNIST Dataset on Raspberry Pi}
\label{tab:my_label}
\begin{tabular}{|p{3.0cm}|p{2.5cm}|}
\hline
\textbf{Attack Strength} & \textbf{Efficacy ($\%$)}  \\
\hline
0.01 & 2.5\\ \hline
0.03 & 4.3\\ \hline
0.08 & 16.3 \\ \hline
0.1 & 24.5\\ \hline
0.31 & 67.9\\ \hline
0.35 & 67.6\\ \hline
0.4 & 67.6\\\hline
\end{tabular}
\end{table}
\par 

\subsection{Implementing Gesture Recognition on ESP32}
In this section, our emphasis is on an in-depth analysis into the susceptibility of AI models deployed on the ESP32 platform, coupled with a detailed summary of the quantitative performance metrics observed. Our experimental methodology involved the systematic execution of model extraction and evasion attack experiments, with an emphasis on a Gesture Recognition Dataset. The objective was to assess the ESP32's computational robustness in precisely recognizing and classifying gestures, particularly when subjected to adversarial perturbations.
\subsubsection{Model Extraction Attack}
As we observed that on the host platform, we successfully replicated the functionalities of the original model with an accuracy of 69.23 $\%$. Subsequently, we compiled the model into the TensorFlow Lite (TFLite) format and further transformed it into Bytes format to facilitate its deployment onto the ESP32 device. This conversion ensured compatibility and optimization for the execution of our AI model on the ESP32.
Upon conducting inference on the ESP32 using the converted Bytes model, a noteworthy observation was made: the model's accuracy remained consistent, maintaining its previous accuracy level of 69.23 $\%$. This signifies that the deployment procedure, encompassing the conversion of the model into a compatible format for the ESP32, did not introduce adverse effects on its performance. However, it is important to acknowledge that the model also remains susceptible to adversarial attacks.

\renewcommand{\arraystretch}{1.5}
\begin{table}[t]
  \centering
  \label{tab:my_label}
    \caption{Result of Gesture Recognition Dataset Model Extraction Attack on ESP32}
  \begin{tabular}{|c|c|c|}
    \hline
  \textbf{Model} &  Victim Model (\textit{V}) & Surrogate Model (\textit{S})\\ 
    \hline
{\textbf{Accuracy ($\%$)}} &  92.37 & 69.23  \\ \hline
\textbf{Model Format} & \multicolumn{2}{|c|}{Bytes}  \\ \hline \cline{2-3} 
\textbf{Model Size (MB)} &  0.0839 & 0.0839   \\ \hline

  \end{tabular}
\end{table}
\par Table VI provides an overview of the results from the model extraction attack, presenting details of the deployed model on ESP32. Furthermore, it offers a comparative assessment of accuracy and model size, drawing comparisons between the \textit{V} and the \textit{S}.
\renewcommand{\arraystretch}{1.5}
\begin{table}[t]
\centering
\caption{Result of Attack Strength vs Efficacy for Gesture Recognition Dataset on ESP32}
\label{tab:my_label}
\begin{tabular}{|p{3.0cm}|p{2.5cm}|}
\hline
\textbf{Attack Strength} & \textbf{Efficacy ($\%$)}  \\
\hline
0.01 & 12.17 \\ \hline
0.05 & 22.03\\ \hline
0.1 & 31.36 \\ \hline
0.3 & 53.39\\ \hline
0.5 & 71.87\\ \hline
0.7 & 86.44\\ \hline
0.9 & 95.76\\\hline
\end{tabular}
\end{table}
\subsubsection{Evasion Attack}
In this section, we'll evaluate the model's robustness in the evasion attack on the ESP32 platform following the conclusion of the model extraction phase. Our primary objective was to assess the robustness and susceptibility of the \textit{V} to evasion attacks when deployed on the ESP32.
The outcomes and significant findings derived from these experiments have been summarized and presented in Table VII. This table serves as a comprehensive representation of the experiment's results, providing a clear overview of the model's performance under different attack strengths. It highlights any instances of misclassification or successful evasion by the generated adversarial examples, enabling a concise analysis of the model's vulnerabilities.

\section{Potential Defense}
From the above experiments, we saw that how attacks crafted on a powerful host machine can be transferred seamlessly to a non-secure Tiny devices. AI defense mechanisms are critical components in safeguarding AI models deployed across various systems and devices. These defenses encompass a range of techniques and strategies designed to protect AI models from adversarial attacks, data breaches, and unauthorized access. Simple hardware defenses involve basic measures such as encryption, access controls, and secure boot mechanisms. While these methods provide a foundational level of security, they may not be sufficient for protecting AI models deployed on resource-constrained devices. Complex hardware defense methods, such as secure communication protocols and hardware-based intrusion detection systems, offer more advanced protection against security threats. However, implementing these methods on low-power devices presents significant challenges. Low-power devices prioritize energy efficiency and may not have the necessary hardware components or processing capabilities to support complex security protocols. Resource-constrained devices, such as IoT sensors, edge computing devices, and embedded systems, often have limited processing power and memory. As a result, they may lack the computational resources needed to implement robust security measures at the hardware level.
\section{Conclusion}
Our experimentation reveals that the vulnerability of AI models to adversarial attacks remains consistent across various deployment scenarios, whether on cloud-based servers or edge devices. Factors like hardware type or model size have minimal impact on this risk. Safeguarding AI models on edge devices is paramount as edge computing gains prominence. Prioritizing defense against adversarial threats ensures the reliability, security, and trustworthiness of AI-powered systems, regardless of industry or organization. 
As part of our future work, we intend to conduct more extensive experiments on a broader range of low-power, resource-constrained devices. This comprehensive approach will provide us with the opportunity to rigorously evaluate the effectiveness of diverse defense mechanisms against adversarial attacks. This empirical understanding will facilitate the development of tailored security solutions precisely calibrated to the distinct characteristics and limitations of individual hardware configurations. This fine-tuning process aims to enhance the overall robustness and resilience of AI-powered systems within real-world deployment scenarios.

\bibliographystyle{IEEEtran}
\bibliography{ref}
\vspace{12pt}

\end{document}